\documentclass[preprint,aps,floats,tightenlines]{revtex4}

\preprint{\vbox{
\hbox{DOE/ER/40762-257}
\hbox{UMD-PP-02-057}
}} \bigskip \bigskip

\begin{document}
\title{Constraints on Two-Body Axial Currents from Reactor Antineutrino-Deuteron
Breakup Reactions}
\author{Malcolm Butler}
\affiliation{Department of Astronomy and Physics, Saint Mary's University\\
Halifax, NS B3H 3C3 Canada\\
{\tt mbutler@ap.stmarys.ca}}
\author{Jiunn-Wei Chen}
\affiliation{Department of Physics, University of Maryland \\
College Park, MD 20742, USA\\
{\tt jwchen@physics.umd.edu}}
\author{Petr Vogel}
\affiliation{Department of Physics, California Institute of Technology\\
Pasadena, California 91125, USA\\
{\tt vogel@citnp10.caltech.edu}}

\begin{abstract}
We discuss how to reduce theoretical uncertainties in the
neutrino-deuteron breakup cross-sections crucial to the Sudbury Neutrino Observatory's efforts
to measure the
solar neutrino flux. In effective field theory, the dominant uncertainties
in all neutrino-deuteron reactions
 can be expressed through a single, common, isovector axial two-body
current parameterized by $L_{1,A}$. After briefly reviewing the status of fixing $L_{1,A}$
experimentally, we present a constraint on $L_{1,A}$ imposed by existing
reactor antineutrino-deuteron breakup data. This constraint alone leads to an uncertainty of 6-7\%
at 7 MeV neutrino energy in the cross-sections relevant to the Sudbury Neutrino
Observatory. However, more significantly for the Sudbury experiment, the 
constraint implies an uncertainty of only 0.7\% in the ratio
of charged to neutral current cross-sections used to verify the existence of neutrino
oscillations, at the same energy.  This is the only 
direct experimental constraint from the two-body system, to date, of the uncertainty 
in these cross-sections.
\end{abstract}

\maketitle

\vfill\eject

\section{Introduction}

Recent results from the Sudbury Neutrino Observatory~(SNO) \cite{SNO}
highlight the importance of a precise determination of neutrino-deuteron
breakup reaction cross-sections. 
The three reactions used by SNO to detect the $^{8}B$ solar flux are 
\[
\begin{array}{llll}
\nu _{e}+d & \rightarrow  & p+p+e^{-} & \qquad \text{(CC)}, \\ 
\nu _{x}+d & \rightarrow  & p+n+\nu _{x} & \qquad \text{(NC)}, \\ 
\nu _{x}+e^{-}\!\!\!\! & \rightarrow  & \nu _{x}+\text{{}}e^{-} & \qquad 
\text{(ES).}
\end{array}
\]
\ The charged current reaction (CC) is sensitive exclusively to electron-type
neutrinos, while the neutral current reaction (NC) is equally sensitive to
all active neutrino flavors ($x=e,\mu ,\tau $). The elastic scattering
reaction (ES) is sensitive to all active flavors as well, but with reduced
sensitivity to $\nu _{\mu }$ and $\nu _{\tau }$. Detection of these three
reactions allows SNO to determine the electron and non-electron active
neutrino components of the solar flux, and it is then obvious that the cross sections
for these three reactions are important inputs for SNO. However, while the
ES cross section is very well determined, the CC and
NC cross sections have never been tested to high precision.

Theoretically, the complications in the CC and NC processes are due to
two-body currents which are irreducible interactions involving leptonic
external currents and two nucleons. The two-body currents can be
calculated either through meson exchange diagrams aided by modeling of any
unknown weak couplings, or can be parameterized using effective field theory
(EFT). In both cases, experimental data from some other  process are required
in order to calibrate the unknowns in the problem. In EFT, this calibration procedure
can be described in an economic and systematic way. The reason is that, up to
next-to-next-to-leading order (NNLO) in EFT, all low-energy weak
interaction deuteron breakup processes depend on a common isovector axial
two-body current, parameterized by $L_{1,A}$ \cite{BCK}. This implies that
a measurement of any 
one of the breakup processes could be used to fix $L_{1,A}$.

In this paper, we will briefly review the EFT approach and discuss
experiments that could be used to fix $L_{1,A}$. Then we present the
constraint on $L_{1,A}$ using reactor $\overline{\nu }d$ scattering.

\section{Effective Field Theory}

For the deuteron breakup processes used to detect solar neutrinos, where $E_{\nu}<15$~MeV,
the typical momentum
scales in the problem are much smaller than the pion mass $m_{\pi }(\simeq
140$ MeV$).$ In these systems pions do not need to be treated as dynamical
particles since they only propagate over distances $\sim 1/m_{\pi }$, much
shorter than the scale set by the typical momentum of the problem. Thus the
pionless nuclear effective field theory, ${\rm EFT}({\pi \hskip-0.6em/})$ 
\cite{KSW96,K97,vK97,Cohen97,BHvK1,CRS}, is applicable.

In ${\rm EFT}({\pi \hskip-0.6em/})$, the dynamical degrees of freedom are
nucleons and non-hadronic external currents. Massive hadronic excitations
such as pions and the delta resonance
are ``integrated out,'' resulting in contact
interactions between nucleons. Nucleon-nucleon interactions are calculated
perturbatively with the small expansion parameter 
\begin{equation}
Q\equiv \frac{(1/a,\gamma ,p)}{\Lambda }
\end{equation}
which is the ratio of the light to heavy scales. The light scales include
the inverse S-wave nucleon-nucleon scattering length $1/a(\lesssim 12$ MeV$)$
in the $^{1}S_{0}$ channel, the deuteron binding momentum $\gamma (=45.7$
MeV)  
in the $^{3}S_{1}$ channel, and the typical nucleon momentum $p$ in the
center-of-mass frame. The heavy scale $\Lambda $ is set by the pion mass $
m_{\pi }$. This formalism has been applied successfully to many processes
involving the deuteron~\cite{CRS,npdgam2}, including Compton scattering \cite
{dEFT,GR}, $np\rightarrow d\gamma $ for big-bang nucleosynthesis \cite
{npdgam1,Rupak}, $\nu d$ scattering  for SNO physics \cite{BCK}, the solar $pp$
fusion process \cite{KR,pp}, and parity violating observables \cite{PV}.
Also studies on three-nucleon systems \cite{BHvK2} have revealed 
highly non-trivial renormalizations associated with three body forces in the
$s_{1/2}$ channel (e.g., $^3$He and the triton). For other channels,
precision calculations were carried out to higher orders
\cite{BHvK1,3bodyothers}. 

In Ref.~\cite{BCK}, ${\rm EFT}({\pi \hskip-0.6em/})$
is applied to compute
the cross-sections for four channels (CC, NC, $\overline{\nu }
_{e}+d\rightarrow e^{+}+n+n$ ($\overline{\nu }$CC) and $\overline{\nu }
_{x}+d\rightarrow \overline{\nu }_{x}+n+p$ ($\overline{\nu }$NC)) to
next-to-next-to-leading order (NNLO), up to 20 MeV (anti)neutrino energies.
As already mentioned, these processes have been shown to depend on only 
one parameter, $L_{1,A}$,
an isovector axial two-body current. This dependence is 
subject to an intrinsic uncertainty in our EFT
calculation at NNLO of less than 3\%. Through varying 
$L_{1,A}$, the potential model results of Refs.~\cite{YHH}
and \cite{NSGK} are reproduced to high accuracy for all four channels. 
This confirms that the 
$\sim 5\%$ difference between Refs.~\cite{YHH} and \cite{NSGK} is due largely 
to different assumptions made about short distance physics.

The same two-body current $L_{1,A}$ also contributes to the proton-proton
fusion process $p+p\rightarrow d+e^{+}+\nu _{e}$.  This is the primary reaction
in the $pp$ chain of nuclear reactions that power the sun, reactions which in 
turn generate the neutrino flux to be observed by SNO. The calculations in 
${\rm EFT}({\pi \hskip-0.6em/})$ were carried out first to second order 
\cite{KR}, and then
to fifth order \cite{pp}. 

\section{Fixing $L_{1,A}$}

In order to determine neutrino-deuteron breakup reaction cross-sections and the $pp$ fusion
amplitude to high precision, one needs a precise determination
of $L_{1,A}$. Naive dimensional analysis gives $
\left| L_{1,A}\right| $ $\sim 6$ fm$^{3}$ when the renormalization scale $
\mu $ is set to $m_{\pi }$. This implies that the contribution from $L_{1,A}$ is at
the 7--8\% level at a neutrino energy of 7~MeV. However, this is a $\mu$-dependent 
statement and cannot be used to draw comparisons between two-body contributions in
the EFT calculation and conventional potential model calculations.
It is therefore important to consider how one can constrain $L_{1,A}$
experimentally.  This is a daunting task, as most weak processes where $L_{1,A}$
contributes are difficult (if not practically impossible) to measure 
accurately in the
laboratory. Further, some experiments such as the measurement of the flux-averaged CC 
cross-section using neutrinos
from stopped muon decays~\cite{willis}, employ neutrino energies greater than 20~MeV. At this
time, the convergence
of the calculations of Ref.~\cite{BCK} is uncertain at these higher energies, so we are
forced to ignore such experiments. Here we list some observables that could be used to
determine $L_{1,A}$:

\begin{enumerate}
\item  $\nu _{e}+d\rightarrow e^{-}+p+p$: the planned ORLaND detector \cite
{ORLaND} has proposed to measure this CC process with $\sim 1\%$ accuracy.
This measurement, combined with higher order calculations in EFT, would
calibrate SNO's CC and NC processes to the same level of accuracy. 

\item  reactor $\overline{\nu}_{e}d$ breakup reactions: the main topic of this
letter. We discuss the extraction of $L_{1,A}$ in next section.

\item  solar $\nu _{x}e$ elastic scattering (ES): 
the three channels available for measuring the solar neutrino flux, ES, CC, and NC 
are all measured by SNO, and ES is measured at Super-K to high precision. 
In general, these results can be used to
constrain three quantities: the electron and non-electron active neutrino
components of the solar flux, and $L_{1,A}$. As statistics improve in all three
channels, SNO could become self-calibrating in that it could use this determination
of $L_{1,A}$ to fix the CC and NC cross-sections~\cite{nueelastic}.

\item  muonic atom capture $\mu ^{-}+d\rightarrow \nu _{\mu }+n+n$ \cite
{mumyrer,muchen}$:$ this reaction can involve significant energy and momentum
transfers. ${\rm EFT}({\pi \hskip-0.6em/})$ may fail in 
regions of phase space where the neutrons are energetic, but should be rapidly convergent in the
region of phase space where the two neutrons move slowly~\cite{muchen}. 
Coincident measurements of those slow neutrons to the
required precision are very difficult. However, indirect high-precision
measurements of the total capture rate are feasible at PSI by comparing 
the lifetime of $\mu^-$ to that of $\mu^+$ in a deuterium target
(thereby avoiding the need to detect final-state neutrons), then 
subtracting the (easier to measure) faster neutron events \cite{Kammel}.

\item  Tritium beta decay $^{3}H\rightarrow \,^{3}He+e^{+}+\nu _{e}$: under
the assumption that three-body currents are negligible, Schiavilla et al. 
\cite{Schiavilla} used this process to fix the two-body current and made a prediction 
for the solar fusion process $p+p\rightarrow \,d+e^{+}+\nu _{e}$, with an
accuracy of better than 1\%. This prediction by Schiavilla et al.
\cite{Schiavilla} can be translated to a constraint on $L_{1,A}$ using the
EFT formula of \cite{pp}. After updating the pion-nucleon coupling $g_{A}$ 
from 1.26 to 1.267,
we obtain $L_{1,A}^{\text{Schiavilla et al.}}=6.5\pm 2.4$ fm$^{3}$ at NNNLO. This result
includes the theoretical uncertainty discussed in
Ref.~\cite{pp}. If we truncate the results from Ref.~\cite{pp} at NNLO, the
expression for $\Lambda(0)$ becomes
\begin{equation}
\Lambda_{NNLO}(0)=2.61+0.0104 \bigg({L_{1,A}\over 1\ {\rm fm}^3}\bigg)+{\cal O}(1.5\%)
\label{eqlambdannlo}
\end{equation}
If we compare this to Schiavialla et al., we find
$L_{1,A}^{\text{Schiavilla et al.}}=4.2\pm 3.7$ fm$^{3}$. This latter result is
more appropriate for comparison to the constraints presented later in this work.
We note also that, following the approach of Schiavilla et al.,
other calculations have made use of tritium beta decay to make
predictions for other weak processes, and all claim an accuracy of 
$\sim 1\%$ \cite{t3appli}.

\item  Helioseismology: the ability of the standard solar model to reproduce
acoustic mode ($p$-mode) oscillations in the Sun to high precision 
can be used to constrain the $pp$ fusion cross-section, and thus $L_{1,A}$. 
Assuming that the solar model must reproduce these oscillations to the
same precision, it is found that $L_{1,A}=7.0\pm5.9\ {\rm fm}^3$~\cite{helio} at NNNLO,
or $L_{1,A}=4.8\pm 6.7\ {\rm fm}^3$ using eq.~\ref{eqlambdannlo} at NNLO. These
numbers also include the theoretical uncertainty in the reduced matrix element for
$pp$ fusion \cite{pp} (but not from other theoretical uncertainties in the solar model). It
should be noted that this constraint is more an indicator of the scale and sign
of $L_{1,A}$.  There are other physical inputs to the solar model (e.g., opacities)
that will weaken this constraint if also allowed to vary.

\end{enumerate}

So far, we have only given numerical constraints on $L_{1,A}$ in methods 5 and 6.
These constraints rely on certain assumptions that require deeper theoretical study.  In the 
next section we will look at fixing $L_{1,A}$ using method 2, namely reactor
$\overline{\nu}_ed$ breakup reactions.

\section{A Constraint from Reactor Antineutrino Experiments}

The charged
current $\overline{\nu }_{e}d\rightarrow e^{+}nn$ ($\overline{\nu }$CC) and
neutral current $\overline{\nu }_{x}d\rightarrow \overline{\nu }_{x}np$ ($
\overline{\nu }$NC) deuteron disintegration have been observed in several
experiments with reactor antineutrinos. We can  use the results of these
experiments to constrain the parameter $L_{1,A}$.

The results of the pioneering experiment at 
Savannah River  \cite{Savannah,Reines1} have been subsequently 
revised \cite{Reines2}.
The fuel composition for that
experiment has not been published, and the effect of the
revision on the error bars is uncertain. Thus we do not use the
Savannah River experiment in our fit, but concentrate instead on the more
recent measurements at  Rovno \cite{Rovno}, Krasnoyarsk \cite{Krasnoyarsk}, 
and Bugey \cite{Bugey} where sufficient information
is available. 

The thresholds $(E_{th})$ of the $\overline{\nu }$NC and $\overline\nu$CC reactions 
are 2.23 MeV, and 4.03 MeV, respectively. These relatively high
thresholds, particularly for the $\overline{\nu }$CC reaction, make
the yield more dependent on the reactor fuel composition than
for the usual $\bar{\nu}_e p \rightarrow e^+ n$ reaction. If the fuel
composition is known (i.e., the fraction of fissions corresponding to 
$^{235}$U, $^{238}$U, $^{239}$Pu, and  $^{241}$Pu), one
can evaluate the $\bar{\nu}_e$ flux ${N_{\overline{\nu }}(E)}$ 
in units of the number of $\overline{\nu }_{e}$ per fission, using
the known  $\overline{\nu }_{e}$ flux produced by
each reactor fuel \cite{flux}.  

The results of Refs.~\cite{Rovno,Krasnoyarsk} are expressed as
the averaged cross section (in cm$^{2}/$fission)
\begin{equation}
\overline{\sigma }_{fission}={\int_{E_{th}}^{E_{max}}
N_{\overline{\nu }}(E)\sigma (E)dE,}  \label{sigfiss}
\end{equation}
where $E_{max}$ is the maximum energy available in the reactor spectrum
(the flux ${N_{\overline{\nu }}(E)} \rightarrow 0$ at  $E_{max}$).

The results of Ref.~\cite{Bugey} have a different
normalization (in cm$^{2}$/$\overline{\nu }_{e}$): 
\begin{equation}
\overline{\sigma }_{\overline{\nu }}={\frac{\int_{E_{th}}^{E_{max}}N_{
\overline{\nu }}(E)\sigma (E)dE}{\int_{E_{th}}^{E_{max}}N_{\overline{\nu }
}(E)dE},}
\end{equation}
which can be
converted to $\overline{\sigma }_{fission}$ easily
(Note the erratum to that reference, however). The
measured averaged cross sections $\overline{\sigma }_{fission}$ for each 
of Refs.~\cite{Rovno,Krasnoyarsk,Bugey} are listed in Table I in units of
cm$^{2}/$fission.

The $\overline{\nu }$CC and $\overline{\nu }$NC cross sections were
calculated to NNLO in ${\rm EFT}({\pi \hskip-0.6em/})$ and were parameterized
in the form 
\begin{equation}
\sigma (E)=a(E)+b(E)L_{1,A}\text{,}  \label{cs}
\end{equation}
with the computed $a(E)$ and $b(E)$ tabulated in Ref.~\cite{BCK}. It has been
noted that these cross-sections must be updated to include the most recent
determination of $g_A$ and to
include electromagnetic radiative corrections~\cite{BP}. To change
the pion-nucleon coupling from $g_{A}=1.26$ used in Ref.~\cite{BCK} to the
most up-to-date value $1.267\pm 0.004$
\cite{PDG}, we multiply $a(E)$ by $(1.267/1.26)^{2}$
and multiply $b(E)$ by $(1.267/1.26).$ The inclusion of the electromagnetic
radiative corrections increase $\overline{\nu }$NC by 1.5\% and $
\overline{\nu }$CC by 3.5\% \cite{Kurylov}, based on the method of 
Refs.~\cite{Towner,EM}. Performing the corresponding integrals of 
$a(E)N_{\overline{\nu }}(E)$ and $b(E)N_{\overline{\nu }}(E)$
and comparing with the experiment, we extract the parameter
$L_{1,A}$ from each measurement and it is shown in the last column
of Table I. The central value of $L_{1,A}$ varies dramatically between measurements.
This is largely due to the fact that the term involving $L_{1,A}$ is a small contribution
to the cross-section, in that a change in $L_{1,A}$ of 1~fm$^3$ would produce an ${\cal O}(1\%)$
change in the cross-sections.  The error bars listed are derived from the experimental
uncertainties; they are dominated by the statistics of each measurement which we assume 
(for simplicity) to
be gaussian in our analysis.
Even though
some common systematic errors are present we shall treat each of the three
 $\overline{\nu }$CC and $\overline{\nu }$NC as an independent
determination of $L_{1,A}$ to obtain $\overline{L}_{1,A} = 3.6 \pm 7.1$ fm$^3$
and $3.5 \pm 6.0$ fm$^3$, 
respectively. The grand average of all six experiments
results in $\overline{L}_{1,A} = 3.6 \pm 4.6$ fm$^3$.
These averages were obtained by summing over the individual results
$(L_{1,A})_i$ with the error bar $\delta_i$ 
\begin{equation}
\overline{L}_{1,A} = \sum_i \frac{(L_{1,A})_i}{\delta_i^2}/
\sum_i \frac{1}{\delta_i^2}~,
\end{equation}
with the total uncertainty 
\begin{equation}
\delta[\overline{L}_{1,A}]  = \left(\sum_i\frac{1}{\delta_i^2}\right)^{-1/2}
\end{equation}
Clearly, the averaging resulted in a substantial reduction of the uncertainty.
For this to be reasonable, the measurements must be truly independent.
The total $\chi^2$ of the averaging procedure is only $\chi_{tot}^2 = 2$.
This relatively (but not alarmingly) low value might
suggest that there could be correlations between the 
$\overline{\nu }$CC and $\overline{\nu }$NC results of each set of
experiments. Possible sources of these correlations include: the simultaneous extraction of $\overline{\nu }$CC and
$\overline{\nu }$NC cross-sections through the
measurement of two 
and one neutron events for $\overline{\nu }$CC and, respectively,
$\overline{\nu }$NC, and; the dependence on the calculated antineutrino spectra (though
this is a small effect). It is difficult to speculate on other possible sources of correlations
in the data.
 
\begin{table}
\begin{tabular}{|c|c|c|}
\hline
& $\overline{\sigma }_{fission}(10^{-44}cm^{2}/fission)$ & $L_{1,A}(fm^{3})$
\\ \hline
$\overline{\nu }$CC$_{\text{ Rovno}}$ & $1.17\pm 0.16$ & $17.4\pm 13.9$ \\ 
\hline
$\overline{\nu }$NC$_{\text{ Rovno}}$ & $2.71\pm 0.47$ & $-2.0\pm 13.8$ \\ 
\hline
$\overline{\nu }$CC$_{\text{ Krasnoyarsk}}$ & $1.05\pm 0.12$ & $-1.3\pm 9.5$
\\ \hline
$\overline{\nu }$NC$_{\text{ Krasnoyarsk}}$ & $3.09\pm 0.30$ & $1.8\pm 8.1$
\\ \hline
$\overline{\nu }$CC$_{\text{ Bugey}}$ & $0.95\pm 0.20$ & $-1.5\pm 17.2$ \\ 
\hline
$\overline{\nu }$NC$_{\text{ Bugey}}$ & $3.15\pm 0.40$ & $11.1\pm 11.7$ \\ 
\hline
\end{tabular}
\caption{Average cross-sections for charged and neutral current $\overline{\nu}_ed$
breakup ($\overline\nu$CC and $\overline\nu$NC, respectively) and the inferred values of $L_{1,A}$ from each
of the three reactor experiments at Rovno~\cite{Rovno}, 
Krasnoyarsk~\cite{Krasnoyarsk}, and Bugey~\cite{Bugey}.}
\end{table}

In addition, there are $\lesssim 3\%$ errors in eq.~(\ref{cs})
due to higher order
corrections. This results in a systematic error of
$\lesssim 3$ fm$^{3}$ on $L_{1,A}$.
Summing up these two uncertainties  in quadrature, we have 
\begin{equation}
\overline{L}_{1,A}^{reactor}=3.6\pm 5.5\text{ }{\rm fm}^{3},
\end{equation}
which is consistent with the constraints from tritium beta decay
and helioseismology,
yet totally independent of both.

Given the importance of this result, one may ask whether a new
high precision experiment of the reactor $\bar{\nu}_e$ induced
deuteron breakup
could result in a substantially more precise determination
of $L_{1,A}$. This seems unlikely, since an improvement 
in accuracy to $\pm 3$ fm$^3$ (for example), equal to the above-stated systematic
error, would require the measurement of the cross section
$\overline{\sigma }_{fission}$ to $\sim$3\% accuracy. Such a measurement
seems rather challenging at the present time.

Finally, it should be noted that all reactor deuteron breakup data were obtained 
in very short baseline experiments. Thus, they are unaffected by 
neutrino oscillations, given current constraints \cite{PDG}.

\section{\protect\bigskip Conclusions}

We have discussed how to reduce the theoretical uncertainties in the $\nu d$
breakup cross-sections crucial for SNO's measurements of the solar neutrino flux. In
effective field theory, the dominant uncertainties can be expressed through
an isovector axial two-body current $L_{1,A}$. We have discussed experiments
that can fix $L_{1,A}$ and have presented a constraint on this quantity
from reactor antineutrino deuteron breakup experiments. 
The reactor antineutrino constraint alone on $L_{1,A}$ can be translated to an uncertainty 
of 6-7\% in the CC and NC reactions used by SNO, at a neutrino energy of 7 MeV. 
The most recent results from SNO verifying the existence of solar neutrino
oscillations~\cite{SNO}, however, rely only on the CC/NC ratio which
is insensitive to $L_{1,A}$ \cite{BCK,YHH}. In fact, the reactor constraint on $L_{1,A}$ implies
an uncertainty of only 0.7\% in the CC/NC ratio at 7 MeV \cite{BCK}.
However, a 6-7\% level of uncertainty in cross-sections would have a 
significant effect on 
extractions of the $^8$B flux. It is encouraging that existing analyses using
the methods discussed in 
Section III produce values of $L_{1,A}$ which are consistent in both sign and magnitude
but, clearly,
more work must be done to constrain $L_{1,A}$ to higher precision.

\ \bigskip 

\begin{acknowledgments}
We thank Hamish Robertson for useful discussions
and Hank Sobel for discussions regarding the Bugey experiment. 
MB is supported by the Natural Sciences
and Engineering Research Council (NSERC) Canada. JWC is supported in part by the U.S.
Dept. of Energy under grant No. DE-FG02-93ER-40762
and PV is supported by the U.S. Dept. of Energy grant No. DE-FG03-88ER-40397.
\end{acknowledgments}

\end{document}